\documentclass[printer]{aa}  

\usepackage{graphicx}
\usepackage{txfonts}
\begin{document}

   \title{The magnetic structure of surges in small-scale emerging flux regions}

   \author{D. MacTaggart
          \inst{1,2,3}
          \and
          S.L. Guglielmino\inst{4}
	\and
	A.L. Haynes \inst{5}
	\and
	R. Simitev \inst{1}
	\and
	F. Zuccarello \inst{4}
          }

   \institute{School of Mathematics and Statistics, University of Glasgow, Glasgow, G12 8QW, Scotland, UK \\
	\email{david.mactaggart@glasgow.ac.uk}
	\and
	White Space Research, Abertay University, Dundee, DD1 1HG, Scotland, UK
         \and
              INAF-Osservatorio Astrofisico di Catania, Via S. Sofia 78, I-95123 Catania, Italia
         \and
	Dipartimento di Fisica e Astronomia - Sezione Astrofisica, Universit\`{a} di Catania, Via S. Sofia 78, I-95123 Catania, Italia
	\and
	School of Mathematics and Statistics, University of St Andrews, St Andrews, Fife, KY16 9SS, Scotland, UK}
   \date{}

 
  \abstract
	{}
   {To investigate the relationship between surges and magnetic reconnection during the emergence of small-scale active regions. In particular, to examine how the large-scale geometry of the magnetic field, shaped by different phases of reconnection, guides the flowing of surges.}
   {We present three flux emergence models. The first model, and the simplest, consists of a region emerging into a horizontal ambient field that is initially parallel to the top of the emerging region. The second model is the same as the first but with an extra smaller emerging region which perturbs the main region. This is added to create a more complex magnetic topology and to test how this complicates the development of surges compared to the first model. The last model has a non-uniform ambient magnetic field to model the effects of emergence near a sunspot field and impose asymmetry on the system through the ambient magnetic field. At each stage, we trace the magnetic topology to identify the locations of reconnection. This allows for field lines to be plotted from different topological regions, highlighting how their geometry affects the development of surges. }
   {In the first model, we identify distinct phases of reconnection. Each phase is associated with a particular geometry for the magnetic field and this determines the paths of the surges. The second model follows a similar pattern to the first but with a more complex magnetic topology and extra eruptions. The third model highlights how an asymmetric ambient field can result in preferred locations for reconnection, subsequently guiding the direction of surges.}
   {Each of the identified phases highlights the close connection between magnetic field geometry, reconnection and the flow of surges. These phases can now be detected observationally and may prove to be key signatures in determining whether or not an emerging region will produce a large-scale (CME-type) eruption.}


   \maketitle
%

\section{Introduction}

Emerging flux regions (EFRs) drive much of the dynamic activity in the solar atmosphere. When the magnetic field from such regions interacts with the pre-existing overlying field, reconnection inevitably ensues. This leads to the formation of directed plasma flows, from surges to coronal jets. Small-scale and ephemeral EFRs have been observed for many years now \citep{harvey73}. Recently, however, they have been viewed with the latest observational facilities at our disposal  \citep[e.g.][]{jiang07,brooks07,guglielmino08,guglielmino10,dominguez14}. These have revealed, in stunning detail, the development of surges in the chromosphere. \cite{guglielmino10} study surges as signatures that accompany small-scale flux emergence within the active region (AR) NOAA 10971. They trace the development of a chromospheric surge in detail and reason that its source is due to magnetic reconnection between the new EFR and the ambient field of the old AR. \cite{dominguez14} also follow the development of a chromospheric surge. Further, they analyze the links between coronal and chromospheric responses in relation to surges. Both studies reveal the distinct filamentary structure of surges in exquisite detail. From a modelling perspective, we are now at the stage of being able to produce detailed numerical simulations for the large-scale structure of small emerging regions ($\sim$30 Mm$^2$) ranging from the top of the solar interior to the low corona \citep[e.g.][]{fan01,magara01,archontis04,fang12,leake13}. Comprehensive reviews of this topic can be found in \cite{fan09a}, \cite{hood12} and \cite{cheung14}. Taking advantage of these recent developments, the main goals of this paper are (a) to investigate in numerical simulations the formation and subsequent evolution of surges embedded within the dynamic reconnecting magnetic field of an EFR, and (b) to analyze the results in light of modern EFR observations.

Surges are low-velocity jets with distinct filamentary structures. Traditionally, they have been thought to occur at the edges of EFRs, have speeds of $\sim$20-30 km~s$^{-1}$ and be due to either reconnection or non-equilibrium (when the pressure within a closed magnetic structure exceeds some threshold) \citep{heyvarts77,priest82}. The hydrodynamic evolution of surges has been studied by \cite{steinolfson79} and \cite{shibata82}. They perform one-dimensional hydrodynamic simulations and classify surges into two categories. The first is the `shock tube' type, where surges are produced directly by the pressure gradient force and consist of matter ejected from the explosion. The second type, or `crest shock' type, is produced by the passage of a shock wave through the chromosphere.  Such models have been successful in predicting surge speeds and densities. However, the assumed progenitors (e.g. bright points) can only be included in initial and boundary conditions in such models. In this paper, we follow the large-scale magnetic development that leads to the formation of surges. We develop models that describe the evolving geometry of an emerging magnetic field. There are two key factors related to surges. The first is that the changing 3D geometry of the magnetic field is linked to various reconnection events - the sources of surges. The second is that the geometry of the field is key to guiding the paths of surges.

Magnetohydrodynamic (MHD) simulations of small ARs have not tended to consider surges in detail. Rather, when studying plasma flows, they have focussed on phenomena such as strong jets \citep[e.g.][]{yokoyama95,yokoyama96,archontis05,moreno13} and spicules \citep[e.g.][]{sykora11}. Coronal jets can involve speeds $\sim$100 km~s$^{-1}$, much faster than those normally quoted for surges. These are typically linked with open coronal magnetic fields, whereas surges are normally associated with field below the corona.

Recently \cite{dmac14} (MH14 hereafter) have produced a flux emergence model for the development of coronal flux ropes. By studying the magnetic topology of the region they are able to identify distinct phases of reconnection associated with the formation of flux ropes. We summarize the results here as they will be useful in interpreting the results of the models in the present work. In MH14, a flux rope is made to emerge into the atmosphere and interacts with a horizontal overlying field. The magnetic fields are aligned so that when they interact they are almost anti-parallel. Initially, the reconnection is steady in the sense that it occurs smoothly at a single separator.  Later, however, there comes a point when the current sheet between the EFR and the overlying ambient field becomes unstable. The single separator bifurcates and plasmoids are ejected from the current sheet. After this, there is a rapid rise upwards and strong coronal jets develop. These have been studied in detail \citep{galsgaard05,archontis05}. As the overlying tension is now broken, the EFR can push upwards into the corona. With a combination of shearing and draining, multiple flux ropes with different magnetic topologies are formed.

Since, in the present work, we are concerned with surges rather than coronal jets, we develop \emph{constrained} flux emergence models. By this we mean that we set up an overlying field that is initially \emph{parallel} to the expanding field of the EFR. This is an efficient method to prevent strong reconnection that could lead to the formation of coronal jets. \cite{dmac11} considers flux emergence into an overlying potential bipolar region. The orientations of the two flux systems are also chosen to minimize strong reconnection. For that study the focus is on flaring activity as seen in observations \citep{zuccarello08}. \cite{leake13} study a similar setup for a flux rope emerging into a magnetic arcade. Their results are focussed towards coronal mass ejections (CMEs) and suggest that flux tube emergence is capable of creating non-current-neutralized stable flux ropes that are possible candidates for CMEs. 

In this paper we will consider models similar to MH14. We shall reverse the overlying magnetic field so that the interacting flux systems are close to parallel, at least initially. We will then follow the evolution of the magnetic topology, which is key to determining how and where surges flow.  To avoid ambiguity and to have a useful operational definition, in this paper we define surges as distinct filamentary structures that can be identified in density and velocity profiles of numerical simulations.

The rest of the paper will be as follows: the next section will describe the basic equations and the numerical setup. This is followed by sections describing the three models presented in this paper. The first contains a single EFR that rises into a horizontal ambient field that is initially parallel to the top of the EFR. Distinct phases of reconnection and surge development are identified and described. The second model is the same as the first but with the addition of a smaller EFR beside the original. This perturbation is included to investigate whether or not the general picture of surge development found in the first model holds in a more complex region. The third model contains a non-uniform ambient field and is used to investigate how the magnetic field can enforce a directional bias on the flowing of surges. The paper ends with a discussion and conclusions.


\section{Model description}

\subsection{Basic equations}

 One of the most successful models that captures the bulk properties of the dynamic solar-atmospheric plasma and magnetic field is compressible MHD. The 3D resistive and compressible MHD equations are solved using a Lagrangian remap scheme \citep{arber01}. In dimensionless form, these are
\[
\frac{{\rm D}\rho}{{\rm D} t} = -\rho\nabla\cdot\mathbf{u},
\]
\[
\frac{{\rm D}\mathbf{u}}{{\rm D} t} = -\frac{1}{\rho}\nabla p + \frac{1}{\rho}(\nabla\times\mathbf{B})\times\mathbf{B}+\frac{1}{\rho}\nabla\cdot\mathbf{T}+\mathbf{g},
\]
\[
\frac{{\rm D}\mathbf{B}}{{\rm D} t} = (\mathbf{B}\cdot\nabla)\mathbf{u} - \mathbf{B}(\nabla\cdot\mathbf{u}) +\eta\nabla^2\mathbf{B},
\]
\[
\frac{{\rm D}\varepsilon}{{\rm D} t} = -\frac{p}{\rho}\nabla\cdot\mathbf{u} + \frac{1}{\rho}\eta|\mathbf{j}|^2 + \frac{1}{\rho}Q_{\rm visc},
\]
\[
\nabla\cdot\mathbf{B} = 0,
\]
with specific energy density
\[
\varepsilon = \frac{p}{(\gamma-1)\rho}.
\]
The basic variables are the density $\rho$, the pressure $p$, the magnetic induction $\mathbf{B}$ (referred to as the magnetic field) and the velocity $\mathbf{u}$. $\mathbf{j}$ is the current density, $\mathbf{g}$ is gravity (uniform in the $z$-direction) and $\gamma =5/3$ is the ratio of specific heats. The dimensionless temperature $T$ can be found from
\[
T = (\gamma-1)\varepsilon.
\]
We make the variables dimensionless against photospheric values, namely, pressure $p_{\rm ph} = 1.4\times 10^4$ Pa; density $\rho_{\rm ph} = 2\times 10^{-4}$ kg~m$^{-3}$; scale height $H_{\rm ph}=170$ km;  surface gravity $g_{\rm ph} = 2.7\times 10^2$ m~s$^{-2}$; speed $u_{\rm ph} = 6.8$ km~s$^{-1}$; time $t_{\rm ph} = 25$ s; magnetic field strength $B_{\rm ph} = 1.3\times 10^3$ G and temperature $T_{\rm ph} = 5.6\times 10^3$ K. In the non-dimensionalization of the temperature we use a gas constant $\mathcal{R}=8.3\times 10^{3}$ m$^2$~s$^{-2}$~K$^{-1}$ and a mean molecular weight $\tilde{\mu}=1$. $\eta$ is the resistivity and we take its value to be $10^{-3}$. This value is close to the lowest \emph{physical} resistivity that can be chosen before \emph{numerical} resistivity dominates \citep[see][]{arber07,leake13}.  The fluid viscosity tensor and the viscous contribution to the energy equation are respectively
\[
\mathbf{T} = \nu\left(e_{ij}-\frac{1}{3}\delta_{ij}u_{i,i}\right),
\]
\[
 Q_{\rm visc} = \nu e_{ij}\left(e_{ij}-\frac{1}{3}\delta_{ij}u_{i,i}\right),
\]   
where $e_{ij}=\frac{1}{2}(u_{i,j}+u_{j,i})$, $\delta_{ij}$ is the kronecker delta and $u_{i,j} \equiv \partial u_i/\partial x_j$. We take $\nu = 10^{-5}$ and use this form of viscosity to aid stability. The code accurately
resolves shocks by using a combination of shock viscosity \citep{wilkins80} and
Van Leer flux limiters \citep{vanleer79}, which add heating terms to the
energy equation. Values will be expressed in non-dimensional form unless explicitly stated otherwise. 

The equations are solved in a Cartesian computational box of
(non-dimensional) sizes, [-130, 130]$\times$[-130, 130]$\times$[-25, 170] in the
$x$, $y$ and $z$ directions respectively. The boundary conditions are
closed on the top and base of the box and periodic on the sides for the first two models presented. For the third model, all the boundaries are closed.  Damping layers are included at the side and top boundaries to reduce the reflection/transmission of waves. The computational mesh contains 372$\times$372$\times$480 points. This resolution is suitable for resolving the evolution of surges.

\subsection{Initial conditions}
The initial idealized equilibrium atmosphere is given by prescribing the temperature profile
\[
T(z) = \left\{\begin{array}{cc}
1-\frac{\gamma-1}{\gamma}z, & z < 0, \\
1, & 0 \le z \le 10, \\
T_{\rm cor}^{[(z-10)/10]}, & 10 < z < 20,\\
T_{\rm cor}, & z \ge 20,
\end{array}\right.
\]
where $T_{\rm cor} = 150$ is the initial coronal temperature. The solar interior is in the region $z<0$, the photosphere and chromosphere lie in $0\le z \le 10$, the transition region occupies $10 < z < 20$ and the corona is in $z \ge 20$. The other state variables, pressure and density, are found by solving the hydrostatic equation in conjunction with the ideal equation of state
\[
\frac{{\rm d}p}{{\rm d}z} = -\rho g, \quad p = \rho T.
\]
The ambient magnetic field will depend on the model in question and will be described in the corresponding sections.

The initial toroidal flux rope, that is placed in the solar interior, has the form

\[
B_x = B_{\theta}(r)\frac{s-s_0}{r},
\]

\[
B_y = -B_{\phi}(r)\frac{z-z_0}{s}-B_{\theta}(r)\frac{xy}{rs},
\]

\[
B_z = B_{\phi}(r)\frac{y}{s} - B_{\theta}(r)\frac{x(z-z_0)}{rs},
\]
with
\[
r^2 = x^2 + (s-s_0)^2, \quad s-s_0 = r\cos\theta, \quad x=r\sin\theta,
\]
and
\[
B_{\phi}=B_0e^{-r^2/r_0^2}, \quad B_{\theta} = \alpha r B_{\phi} = \alpha rB_0e^{-r^2/r_0^2}.
\]
This is derived from a regular expansion of a Grad-Shafranov equation \citep{dmac09}. To leading order, the solutions for a uniformly twisted cylindrical flux tube can be used to describe the magnetic field of a toroidal tube locally. These are then transformed to the global toroidal geometry, giving the required shape for the initial flux tube. The axis of the flux tube is positioned along the $y$-axis. $s_0$ is the major axis of the tube and $r_0$ is the minor axis (of its cross-section). $z_0$ is the base of the computational box. $\alpha$ is the initial twist and $B_0$ is the initial axial field strength. A study of how varying these parameters affects flux emergence is presented in \cite{dmac09}. The values of the parameters will be specified in the following sections. Numerical experiments are also tested in larger domains (e.g. [-160,160]$^2\times$[-25,170]) with the same number of grid points in order to increase confidence in the results.


\section{Single emerging region}

We begin with our simplest flux emergence model - a toroidal flux tube emerging into a horizontal magnetic field. We choose a horizontal ambient magnetic field of 0.01 (13 G). The orientation is chosen so that it is almost parallel to the field
of the emerging flux tube as it expands into the atmosphere. This is along the positive $x$--direction in these simulations. With this choice of ambient field, we can make direct comparisons with MH14. Also a horizontal field is an approximation to a wide overlying arcade. It is a potential field and is simple to incorporate and control in the initial condition. For this model, we adopt the same initial parameter values as MH14. These are the magnetic field strength at the axis, $B_0 = 5$ (6.5 kG), the uniform twist, $\alpha = 0.4$, the major radius (axis height from the base) of the tube, $s_0 = 15$, the minor (cross-sectional) radius, $r_0 = 2.5$ and the base of the domain, $z_0 = -25$. To initiate the experiment, the entire tube is made buoyant. That is, a density deficit relative to the background density is introduced. We shall not describe all details relating to flux emergence here but only focus on activity related to surges. General descriptions of flux emergence can be found elsewhere \citep[e.g.][]{murray06,dmac09,fan09b,hood12}.

\subsection{Phase 1}

As the EFR pushes upwards into the atmosphere, we track the topology of the magnetic field in the same way as described in MH14. Each flux system is given a different colour. We choose the same colour scheme as MH14, the key is: cyan for field lines connecting one side of the simulation box to the other; green for field lines connecting one footpoint of the EFR to the other; blue for field lines connecting one side of the computational domain to one of the EFR footpoints; red for field lines connecting the other side of the computational domain to the other footpoint. 

  \begin{figure}[h]
   \centering
   \includegraphics[width=\hsize]{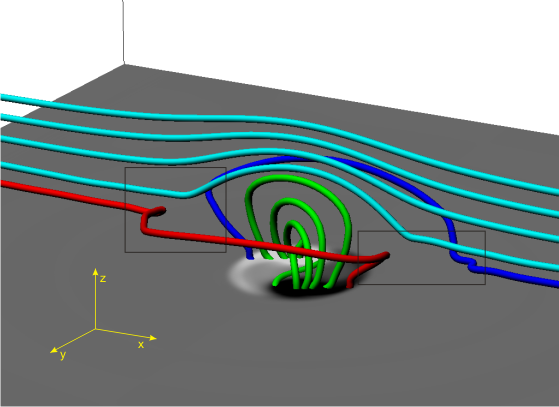}
      \caption{EFR at $t=70$. The different flux regions are colour-coded (see text for details).  The approximate positions of reconnection are indicated by boxes. At these locations, at a height of $z \approx 10$, surges are generated. A map of $B_z$ is displayed at the base of the photosphere ($z=0$) and reveals that the reconnected field lines connect down to the edges of the EFR.}
         \label{edge}
   \end{figure}

\begin{figure}[h]
   \centering
   \includegraphics[width=\hsize]{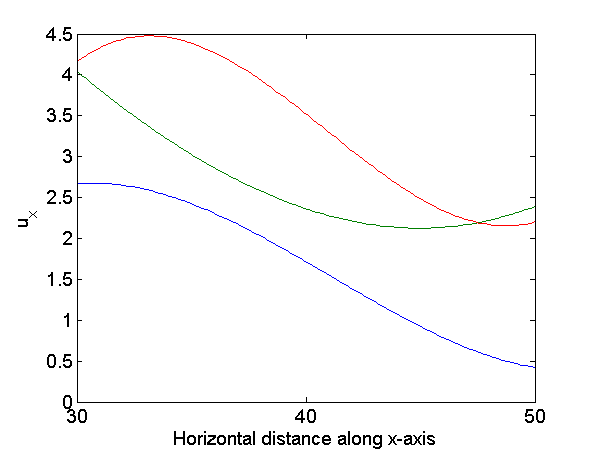}
      \caption{Values of $u_x$ in the $y=0$ plane  at $z=10$, along the surge moving in the positive $x-$direction. Key: blue, $t=64$; green, $t=68$, red, $t=72$.}
         \label{initial_surge_vel}
   \end{figure}

Figure \ref{edge} displays the magnetic field line structure at $t=70$. The EFR is expanding into the atmosphere but there is no overlying reconnection as the two corresponding flux systems are very close to parallel. Reconnection first occurs at the side edges of the EFR. Due to the three dimensionality of the problem, the flux systems will not always be parallel and locations will emerge that are conducive to reconnection. The reconnected (red and blue) field lines kink at  $z\approx 10$ (top of the chromosphere) and connect down to the edges of the EFR. From these kinks (reconnection positions) surges emerge. The surges follow paths away from the EFR and, due to the symmetry of the initial setup, occur on both sides. Figure \ref{initial_surge_vel} shows the development of the velocity profile along the surge. Only one surge is shown as, due to symmetry, the other is the same but in the opposite direction.  The surge is identified by a clear jet-like flow in the velocity data.

The horizontal speeds ($u_x$) are shown at $z=10$ for three different times. These are taken in the $y=0$ plane due to the symmetric setup. $x=30$ is, approximately, the starting position for the surge. This changes in time due to the horizontal expansion of the EFR. The surge itself is almost horizontal due to the geometry of the magnetic field in the model.  At $t=64$ (the blue line) the horizontal speed decreases almost linearly along the surge. The horizontal speed, in dimensional units, at $x=30$ is $u_x= 17$ km~s$^{-1}$. This order of magnitude matches well with typical quoted values for surges \citep[e.g.][]{priest82}. At $t=68$ (the green line) the profile is no longer decreasing monotonically. At $x=30$, the value of the horizontal speed has increased. One must be cautious in interpreting this, however, as a contribution to this value comes from the horizontal expansion of the EFR. This value decreases near $x=40$ which is the outer edge of the EFR just before the start of the surge. The speed then increases at $x=50$ to a value close to that of the start of the surge at $t=64$. The profile at $t=72$ demonstrates the lateral expansion of the EFR.  The peak in the horizontal speed at $x=35$ is due to the reconnection point (the source of the surge) being pushed by the expanding EFR. 

\begin{figure}[h]
   \centering
   \includegraphics[width=\hsize]{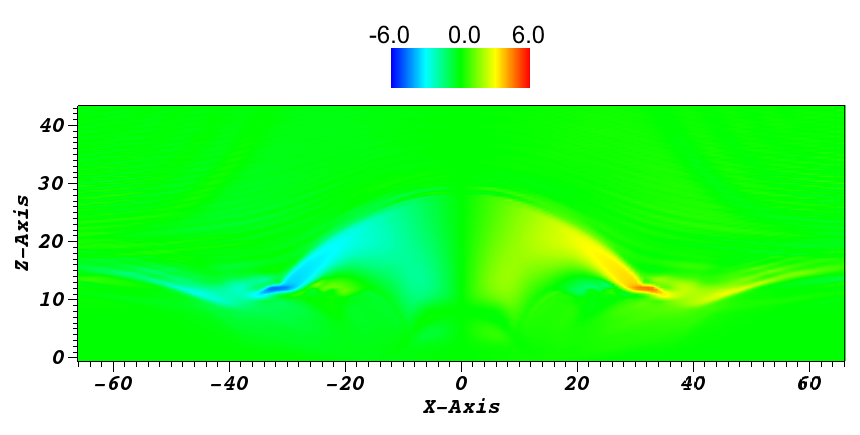}
      \caption{Slice through $y=0$ at $t=68$ showing $u_x$. The EFR expands horizontally and the flow of the surges is shown by the `wings' of the colourmap, beyond $x\approx \pm 30$.}
         \label{slices_surges_t68}
   \end{figure}

Due to the symmetry of the model, a surge with identical properties that travels in the opposite direction appears on the other side of the EFR. This is shown in Figure \ref{slices_surges_t68}, which displays a slice of $u_x$ in the $y=0$ plane at $t=68$. The shape of the velocity map follows that of the magnetic field shown in Figure \ref{edge} with the emerging field pressing into the ambient field. At $(x,z) \approx (\pm 30, 10)$, there is an increase in the horizontal speed due to the positions of reconnection (cf. the positions of the kinks in Figure \ref{edge}). From these points emanate near-horizontal flows.

In short, the surges move horizontally  along the magnetic field with a speed that peaks at $u_x\approx$ 30 km~s$^{-1}$ and exist for several minutes (the first three minutes, measured from the first appearance of the surge, are shown in Figure \ref{initial_surge_vel}). The full lifetime of surges depends on the geometry of the ambient field and the size of the domain. In larger domains where the magnetic field has a similar geometry to that discussed above (i.e. horizontal at the boundaries), surges take longer to reach the boundaries.

\subsection{Phase 2}

\begin{figure}[h]
   \centering
   \includegraphics[width=\hsize]{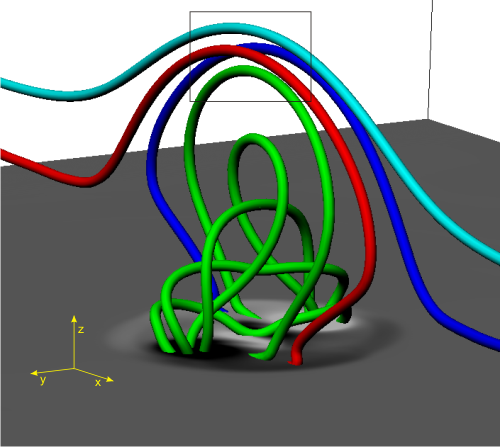}
      \caption{Overlying reconnection at $t=110$. The colour scheme is as before. Reconnection occurs between the top of the (green) EFR and the (cyan) overlying field, within the displayed box. Reconnected (red and blue) field lines allow surges to flow from the corona (top of the EFR) down to the chromosphere (horizontal parts of the reconnected field lines). A  map of $B_z$ is displayed at the base of the photosphere ($z=0$). The non-horizontal parts of the reconnected field lines connect down to the edges of the EFR.}
         \label{mag_field_t110}
   \end{figure}

In Phase 1 the surges occured at the sides of the EFR because this was where reconnection occured. As the twisted EFR pushes into the horizontal ambient field, however, a current sheet develops between the two flux systems and reconnection ensues. This second phase of reconnection occurs smoothly at the apex of the EFR. This behaviour is similar to the single-separator reconnection found in MH14. In the present simulation, however, we were unable to determine if the reconnection occurs at a magnetic separator. The most we can say is that it is an example of 3D reconnection \citep{birn07}. Figure \ref{mag_field_t110} demonstrates the reconnection process at $t=110$.  

\begin{figure}[h]
   \centering
   \includegraphics[width=\hsize]{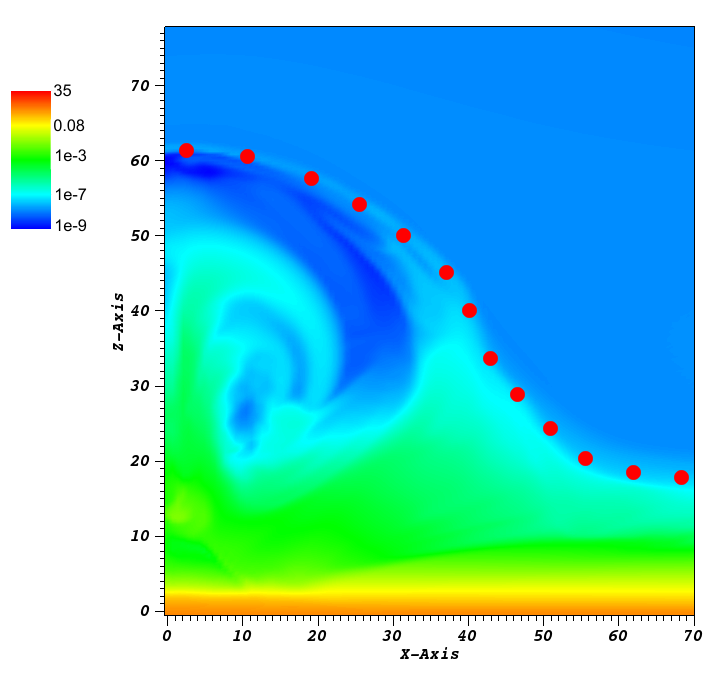}
      \caption{ Slice of $\log\rho$ in the $y=0$ plane at $t=110$. The red dots indicate the approximate edge of the active region.}
         \label{ddots}
   \end{figure}

\begin{figure}[h]
   \centering
   \includegraphics[width=\hsize]{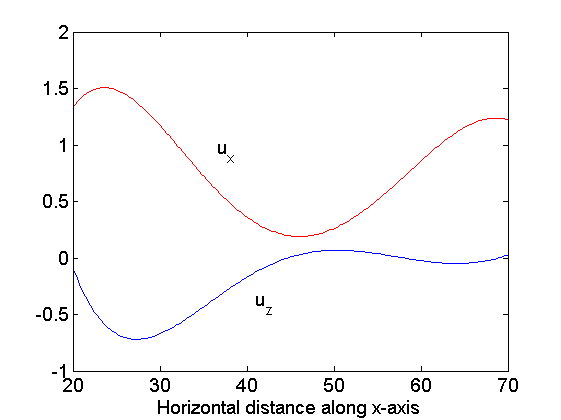}
      \caption{Profiles of $u_x$ and $u_z$ taken in the $y=0$ plane at $t=110$. These represent surge speeds as a function of horizontal distance, evaluated at points on the edge of the EFR in this plane.}
         \label{ux_uz}
   \end{figure}

Reconnection at the top of the EFR (now in the corona) links field lines here down to lower heights, including the chromosphere. This creates a path for surges to flow down to lower heights. By analysing the profiles of $u_x$ and $u_z$ along the paths of reconnected field lines, one can estimate the typical speeds of the surges. Taking various slices through the computational domain at $t=110$, the surges have speeds of $\sim$3 at heights of $z\approx$ 30 to  $\sim$1 at heights below the base of the corona $(z=20)$. In dimensional units, the speeds range from $\sim$20 km~s$^{-1}$ down to $\sim$7 km~s$^{-1}$.  As an example, we shall consider values at the edge of the EFR. Figure \ref{ddots} displays $\log\rho$ in the $y=0$ plane at $t=110$. The approximate edge of the emerging region is highlighted by a path of red circles.  Figure \ref{ux_uz} displays the profiles of $u_x$ and $u_z$ following part of the path shown in Fig. \ref{ddots}. The left-most horizontal position in the figure is near the apex of the EFR and the right-most horizontal position is where the magnetic field is close to horizontal (as shown in Figure \ref{mag_field_t110}). The curve representing $u_x$ shows a nonlinear profile that does not reverse in direction.  The magnitude of the $u_z$ profile has a maximum near the apex of the EFR. This then decreases to approximately zero as the flow becomes close to horizontal.

One important point about this phase of reconnection is that surges travelling horizontally in the chromosphere can be created by reconnection in the corona. \cite{dominguez14} note that they detect a coronal signature before a chromospheric one when studying surges. As the EFR pushes upwards, more field lines are reconnected, i.e. the blue and red flux regions now connect to the corona as well as the chromosphere. This situation continues until a new phase of reconnection sets in.

\subsection{Phase 3}

\begin{figure}[h]
   \centering
   \includegraphics[width=\hsize]{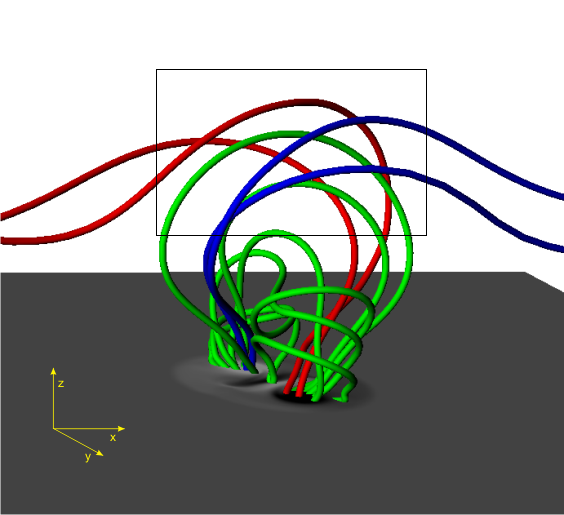}
      \caption{Overlying reconnection at $t=137$. The colour scheme is as before. Reconnection no longer takes place at a single point but at several throughout the current sheet,  as indicated by the box that covers the whole of the top of the EFR. A map of $B_z$ is displayed at the base of the photosphere ($z=0$). The reconnected field lines now connect inside the footpoints rather than at the edges. The ambient field has been omitted for clarity.}
         \label{mag_field_t137}
   \end{figure}

At $t\approx130$, the reconnection no longer occurs smoothly in one location at the top of the EFR. Multiple regions of connectivity form and change rapidly within the current sheet between the EFR and the ambient field. This behaviour is similar to the tearing reconnection found in MH14 in the sense that there is a change from one point of reconnection within the current sheet to several. Figure \ref{mag_field_t137} shows how the reconnected field lines have changed positions, both at the top of the EFR and in connections at the photosphere. Previously, reconnected field lines connected to the edges of the footpoints (cf. Figures \ref{edge} and \ref{mag_field_t110}). Now, however, they connect to the centres of the footpoints.  

\begin{figure}[h]
   \centering
   \includegraphics[scale=0.7]{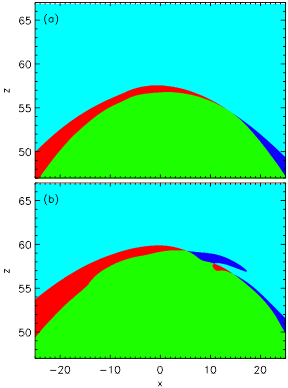}
      \caption{Magnetic connectivity maps. These are for overlying reconnection at (a) $t=136.5$ and (b) $t=137.5$. Both (a) and (b) are on the $y=0$ plane with $(x,z)\in [-25,25]\times[47,67]$. The colour scheme is as before (see text). Moving from (a) to (b), reconnection changes from occuring at a single location, $(x,z)$ = (13.9,~54.7), to three, $(x,z)$ = (5.1,~59.4), (10.6,~57.9) and (14.5,~56.7) . }
         \label{tear}
   \end{figure}

\begin{figure}[h]
   \centering
   \includegraphics[scale=0.47]{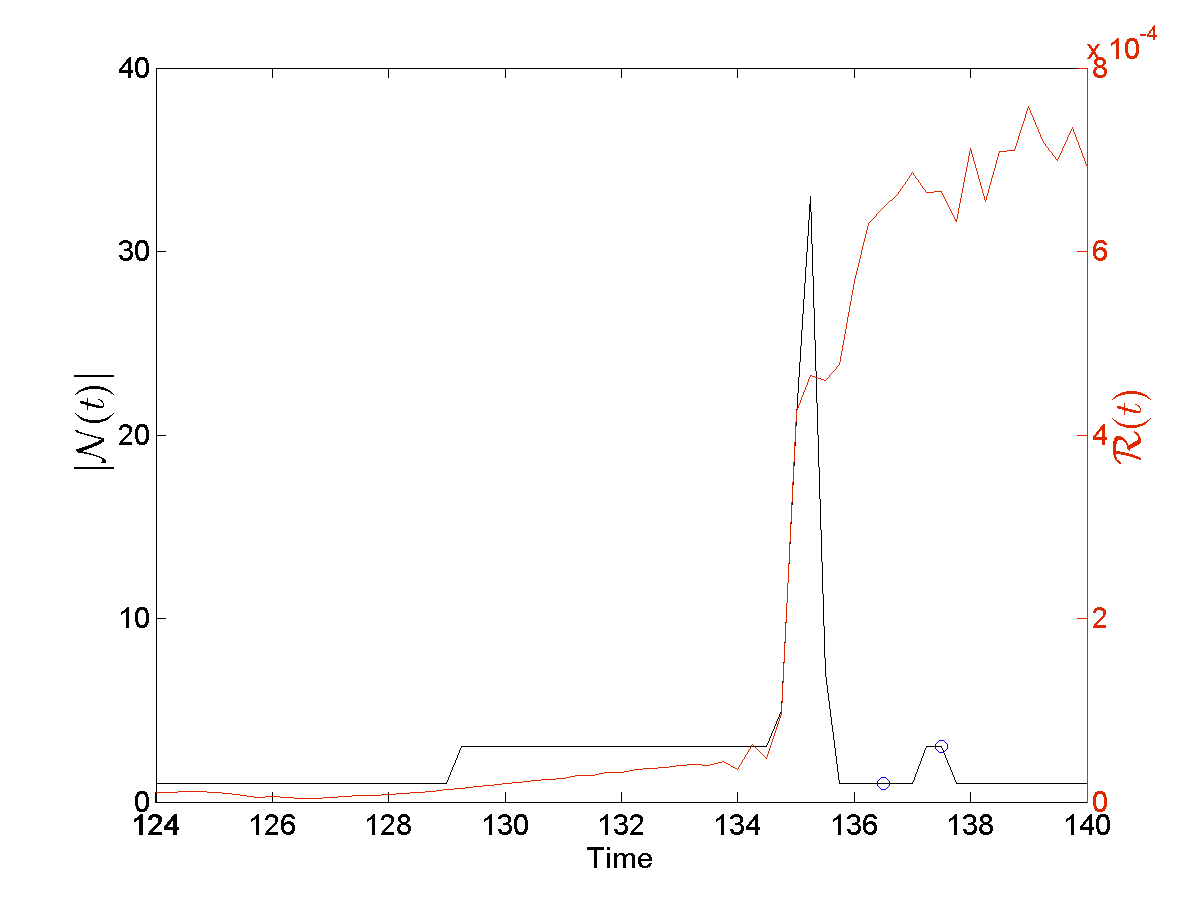}
      \caption{Time series for $|\mathcal{N}(t)|$ (black) and $\mathcal{R}(t)$ (red). Blue circles correspond to the connectivity maps (a) and (b) from Fig. \ref{tear}. The $y=0$ plane has been used as the initial location to trace field lines for the calculation of $\mathcal{R}(t)$.}
         \label{ts}
   \end{figure}

With this change in behaviour, one can draw parallels to previous studies of 3D reconnection. \cite{linton03} present an analysis of 3D reconnection in interacting untwisted flux tubes. In this, initially, simple setup a central point of reconnection forms. This is then disturbed and broken up by the tearing instability. Multiple regions of reconnection emerge and multiple flux tubes are formed. In our simulation, Phase 2 is similar to the initial central reconnection and this degenerates, through a tearing-like instability, into multiple regions of connectivity - Phase 3. This is best shown in connectivity maps, where magnetic field lines, passing through a particular plane, are traced and assigned a colour based on their connectivity. Figure \ref{tear} shows the connectivity maps at the top of the EFR at times $t=136.5$ and $t=137.5$ for the plane $y=0$. The colour scheme follows that of Figures \ref{edge}, \ref{mag_field_t110} and \ref{mag_field_t137}. The maps cover $(x,z)\in [-25,25]\times[47,67]$, the region indicated, approximately, by the box in Figure \ref{mag_field_t137}.  One can see in Figure \ref{tear} (a)  that at $t=136.5$, all four colours (flux regions) meet at a single point $(x,z)$ = (13.9,~54.7). This signature could represent a separator, a quasi-separator or a hyperbolic flux tube. As we have not been able to find separators in this location, this suggests it is one of the latter two choices.  At the later time of $t=137.5$, shown in Fig. \ref{tear} (b), the number of points where the four colours meet increases to three. These are at positions $(x,z)$ = (5.1,~59.4), (10.6,~57.9) and (14.5,~56.7). This means that reconnection is occuring now in several places. If $\mathcal{N}(t)$ is the set of points where all four flux domains meet in the $y=0$ plane at time $t$, Fig. \ref{ts} shows how the number of points $|\mathcal{N}(t)|$ varies with time. To highlight the effect of this on magnetic reconnection, a measure of the reconnection rate $\mathcal{R}(t)$ is also displayed in Fig. \ref{ts}. This is defined by
\[
\mathcal{R}(t) = \max_{\mathcal{N}(t)}\left|\int_{\Gamma}E_{\|}\,{\rm d}l\right|,
\]
where $E_{\|}$ is the parallel electric field integrated along a path $\Gamma$. This path is a field line passing through one of the points in $\mathcal{N}(t)$. The maximum value over all the field lines passing through points in $\mathcal{N}(t)$ is taken as a measure of the reconnection rate. This expression is derived in Appendix A.

During Phase 2,  $|\mathcal{N}(t)|=1$. However, at the onset of Phase 3 just before $t=130$, this value changes to  $|\mathcal{N}(t)|=3.$ It remains steady at this value until $t=134.5$. During this period, $\mathcal{R}(t)$ increases linearly with time. There is then a sudden rise in both $|\mathcal{N}(t)|$ and $\mathcal{R}(t)$ with their rates of change being proportional to each other. $|\mathcal{N}(t)|$ peaks at 33 and then rapidly decreases back to 1. The reconnection rate continues to rise and then fluctuate around 7$\times$10$^{-4}$. Figure \ref{ts} shows that the change from weak reconnection to strong reconnection is accompanied by a complex change in the magnetic topology.

 The change from Phase 2 to Phase 3 is also signaled by high velocities at the top of the EFR. In dimensional units, surges flow through the reconnected field lines with horizontal speeds ranging from  $\sim$14 km~s$^{-1}$ to $\sim$30 km~s$^{-1}$. At later times, the speeds decrease to $\lesssim$ 10 km~s$^{-1}$. The decrease in speeds takes place over a period of $\sim$20 mins.

\subsection{Phase 4}
As well as overlying reconnection, there is the possiblity of internal reconnection within the EFR. In the model of MH14, internal reconnection leads to the formation of a flux rope.{This has also been found in other flux emergence models \citep[e.g.][]{archontis12,leake14}. In the present simulation, internal reconnection also occurs, acting as another source for surges. Figure \ref{mag_field_t200} displays some indicative field lines that reveal the geometry and topology of the magnetic field after internal reconnection has occured at $t=200$.

\begin{figure}[h]
   \centering
   \includegraphics[width=\hsize]{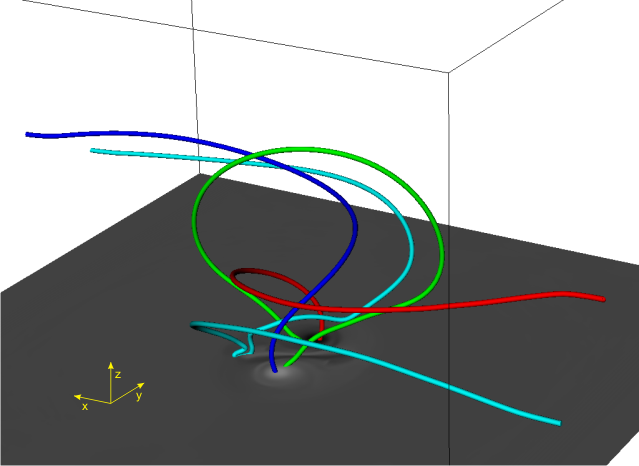}
      \caption{Particular magnetic field lines showing effects of internal reconnection. These are traced to highlight the magnetic topology of the region after internal reconnection has occured at $t=200$. The colour scheme is the same as before.  Previously, cyan field lines represented the overlying ambient field. Here, however, there are also cyan field lines that pass within the EFR, as shown here. A map of $B_z$ is placed at $z=0$.}
         \label{mag_field_t200}
   \end{figure}

The green and red and blue field lines indicate the EFR and the reconnected magnetic flux, respectively, as before. The cyan field line, which connects to both sides of the computational domain, is not part of the ambient magnetic field but is due to internal reconnection. In Phase 3, the red and blue reconnection field lines connect down into the centres of the footpoints. This is also shown in Fig. \ref{mag_field_t200}. Shearing in the EFR brings the red and blue field lines together, causing them to reconnect. This creates magnetic field lines that pass through the EFR. Surges can flow along these field lines and horizontal speeds are detected of $u_x\approx 2$-3. In dimensional units, this is $\sim$ 14 km~s$^{-1}$ to $\sim$21 km~s$^{-1}$.

 \subsection{Summary of phases}
At this point it is helpful to summarize the four stages of reconnection identified in this section. These are collected into the diagram shown in Fig. \ref{cartoon}.
\begin{figure}[h]
   \centering
   \includegraphics[width=\hsize]{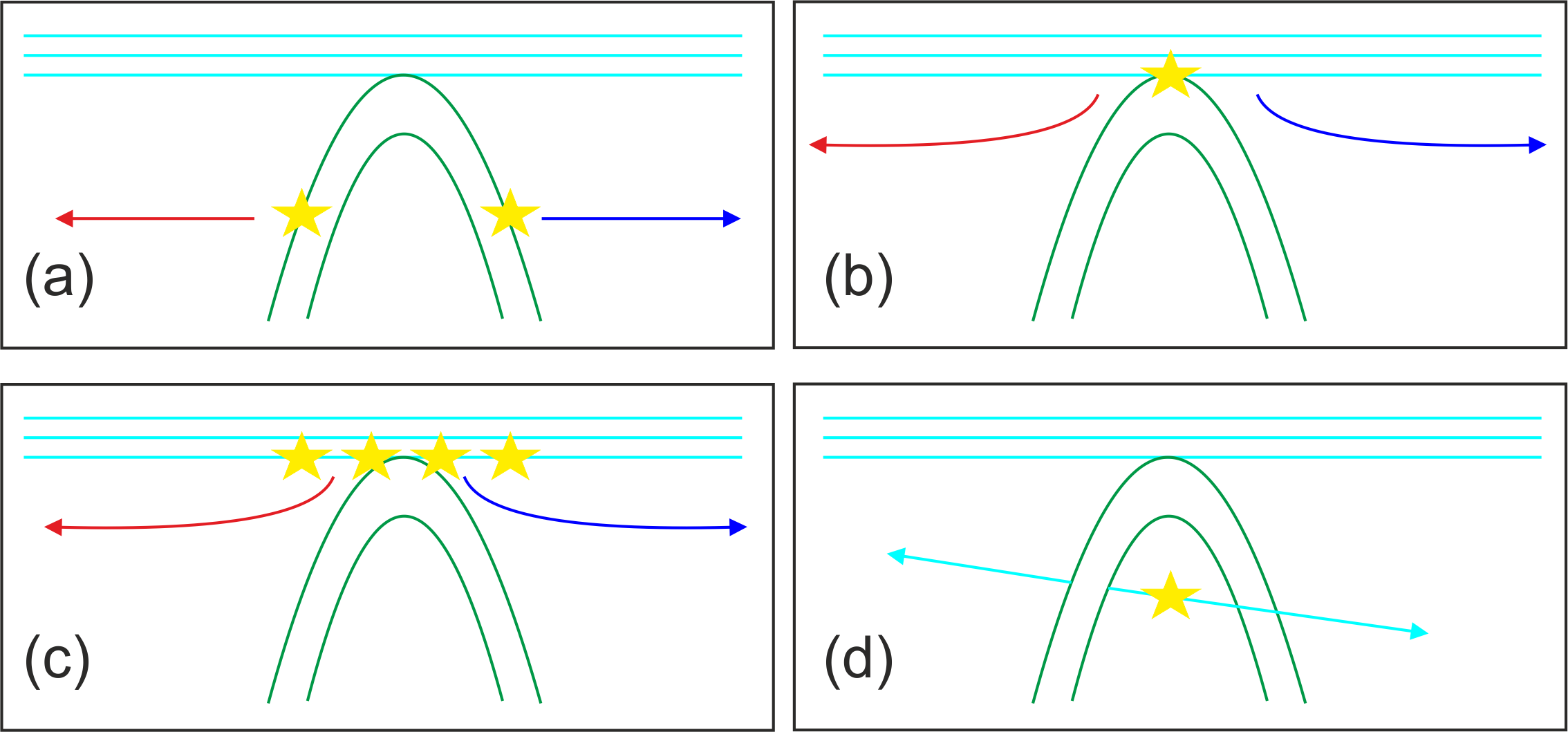}
      \caption{Cartoon of reconnection phases. Each panel highlights a particular phase of reconnection with a simplified representation of the different flux systems. Positions of reconnection are denoted with stars. The colour scheme is the same as previous figures for the first model. Arrows indicate the direction of surges flowing along field lines. (a) is Phase 1, (b) is Phase 2, (c) is Phase 3 and (d) is Phase 4. }
         \label{cartoon}
   \end{figure}
It should be noted that although each panel of Fig. \ref{cartoon} shows a different phase of reconnection, this does not imply that multiple phases cannot occur simultaneously. Panel (a) shows Phase 1, reconnection occuring at the sides of the emerging domain. Panel (b) illustrates how, in Phase 2, reconnection now takes place at the top of the EFR. Eventually, Phase 2 becomes unstable and turns into Phase 3, where there are multiple locations of reconnection at the top of the EFR. This is shown in panel (c). The last panel, (d), displays the internal reconnection of Phase 4. All of these phases are linked to the development and flowing of surges, as has been demonstrated in this section. These phases will reappear in subsequent models for surges and are important for their analysis. 
\section{Perturbed emerging region}

We now consider a more complex version of the emergence model presented in the previous section. Here, we have two emerging regions with the same overlying field as before. There is one main region which has the same parameters as in the previous section. The second region is smaller and acts to perturb the main region. It is placed at $x=-30$ and has the same parameters as the main region with the exception of the initial axial magnetic field being $B_0=-3$ (-3.9 kG). This value has been chosen for two reasons. The first is that its magnitude is smaller than that of the main region. With all else equal, this means that the second region takes longer to emerge. In the present simulation, it begins to interact with the main region at about $t=100$. By this stage, the main region is in Phase 2 of its evolution, as described in the previous section. The second reason has to do with the sign of the axial magnetic field. As our focus is on surges, we do not look for configurations that will encourage strong reconnection. By the time the two EFRs interact one cannot say that the fields will be just parallel, because the 3D geometry is more complicated. However, with this choice of $B_0$ for the second region, the positive  (upward) magnetic field of the second region will press against the positive field of the main region. This is displayed in Fig. \ref{mag_field_t126}, along with selected magnetic field lines at $t=126$. By looking at the map of $B_z$, the positive (white) footpoint of the small region presses against the positive magnetic field of the main region.

\begin{figure}[h]
   \centering
   \includegraphics[width=\hsize]{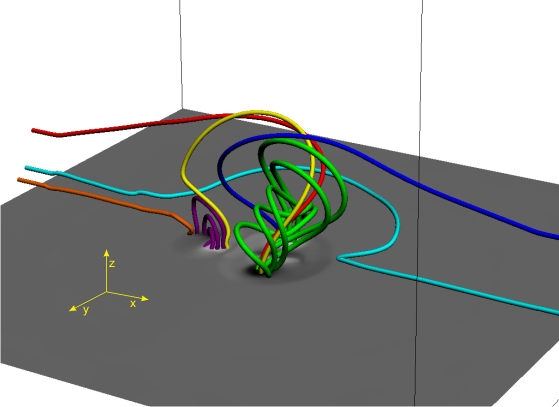}
      \caption{Selected magnetic field lines indicating the magnetic topology at $t=126$. Each colour represents a different flux domain and the key is given in the text. A map of $B_z$ is placed at $z=0$.}
         \label{mag_field_t126}
   \end{figure}

As can be seen from Fig. \ref{mag_field_t126}, the addition of just one new bipole region greatly increases the complexity of the magnetic topology. The colour key is as follows: green connects to both footpoints of the main region; cyan connects to both sides of the computational domain; blue connects from the positive main region footpoint to one side of the domain; red connects the negative footpoint of the main region to the other side of the domain; purple connects both footpoints of the second region; orange connects the negative footpoint of the second region to one end of the computational domain; yellow connects the positive footpoint of the second region to the negative footpoint of the main region. There are other flux domains, the positive footpoint of the second region to one end of the computational domain and the negative footpoint of the second region to the positive footpoint of the main region, that we have not shown for clarity. Those that are shown illustrate different types of behaviour. The cyan field line shows that reconnection occurs at the side of the main region. This generates surges as described in Phase 1 for the previous model.

By $t=126$, the break up of the current sheet has made the overlying reconnection region complex, as in Phase 3 of the previous model. There is now, however, the added complexity of reconnection between the two EFRs. This is shown in Fig. \ref{mag_field_t126} by the yellow field line.  Since this field line passes through the complex reconnecting region at the top of the main EFR (as described for Phase 3), it connects down to the centre of the negative footpoint in the main region. The perturbing region is not in Phase 3, however, and so the yellow field line only connects to the outer edge of the footpoint.

There is also reconnection resulting in field lines connecting the negative footpoint of the second region to the positive footpoint of the main region. This complex mixing of different flux regions results in many reconnection regions that generate surge-like speeds. From $t=100$ to $t=130$ (12.5 mins), typical speeds range from $|\mathbf{u}|\approx 1-4$ ($\sim$7-27 km~s$^{-1}$). Flows within the main EFR are difficult to follow due to the highly dynamic reconnection. Horizontal surges, however, can be detected flowing along the horizontal ambient field, as in the previous model. Figure \ref{density_t126_z19_2} displays a slice showing $\log\rho$ at $z=19$ (just below the corona). The sigmoid of the main EFR can clearly be seen in the centre. To the sides of this, along $y=0$, the surges are identifiable as thin strips of enhanced density. The asymmetry between both sides is due to the perturbation of the second EFR (at approximately (-40,-10) on the slice). This tilts the main EFR, as it pushes into it, and dynamically changes the topology through reconnection, as described above. The horizontal surges develop at various heights, ranging from $z\approx 15 - 40$.

\begin{figure}[h]
   \centering
   \includegraphics[width=\hsize]{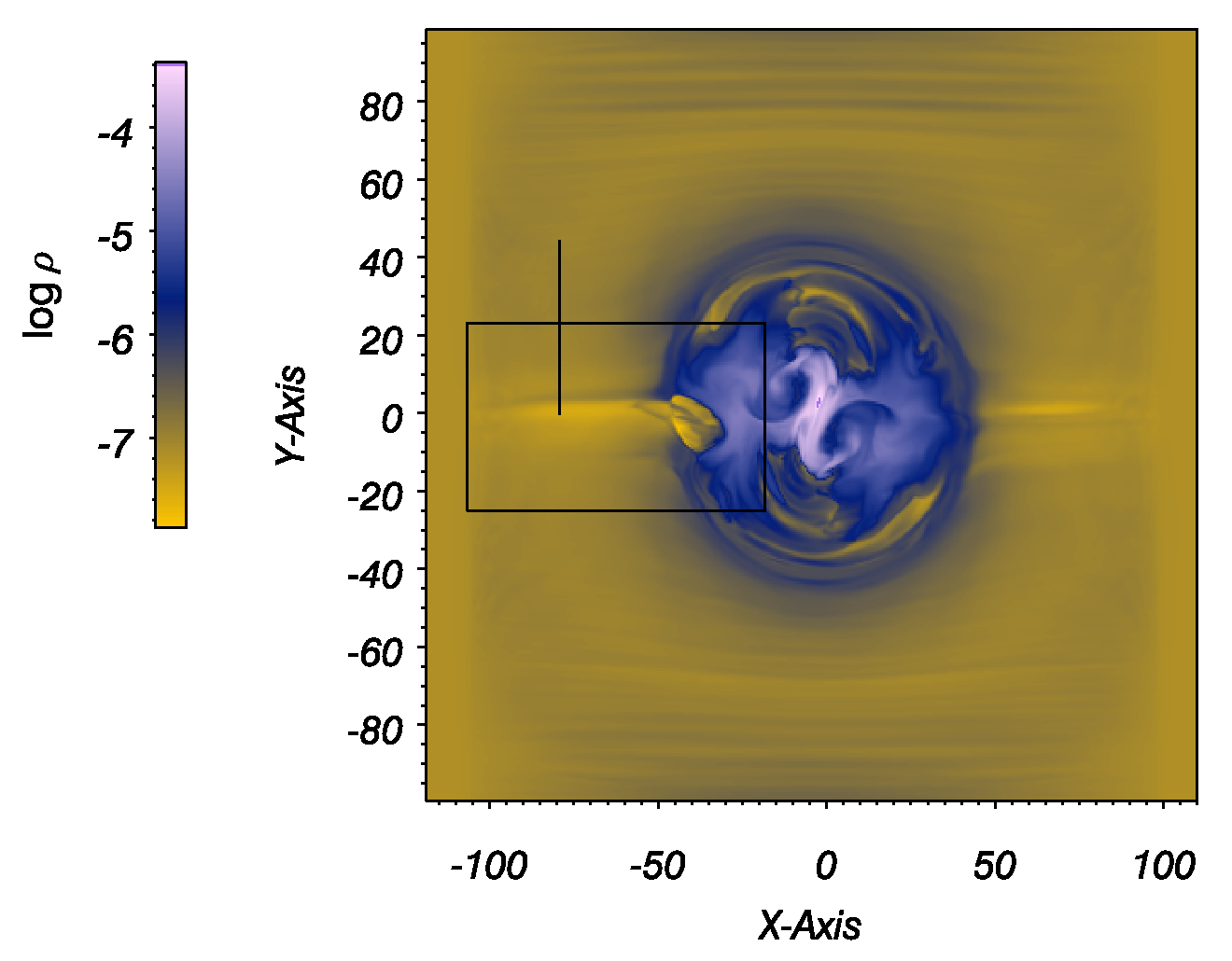}
      \caption{Slice at $z=19$ displaying $\log\rho$ at $t=126$. The sigmoid of the main EFR can be seen at the centre. On either side of this, horizontal surges can clearly be seen. The asymmetry in their formation is due to the presence of the second (perturbative) region. The black rectangle shows the region plotted in Fig. \ref{surge_evol} and the vertical line shows the cut taken for Fig. \ref{surge_den_slice}.}
         \label{density_t126_z19_2}
   \end{figure}

In Fig. \ref{density_t126_z19_2}, there is one clear isolated surge on the left-hand side of the active region. Figures \ref{surge_evol} and \ref{surge_den_slice} show the development of this over the first minute of its evolution since its appearance at $z=19$. Figure \ref{surge_evol} shows the evolution of the density map within the first minute. The region taken for this is highlighted by a black box in Fig. \ref{density_t126_z19_2}.  As can be seen from the maps at different times, there is a distinct jet-like density enhancement corresponding to the surge. Within the time shown, the surge develops much faster than the expansion of the EFR. This shape is highly reminiscent of H$\alpha$ intensity observations of surges \citep[e.g.][]{guglielmino10}.

\begin{figure}[h]
   \centering
   \includegraphics[scale=0.7]{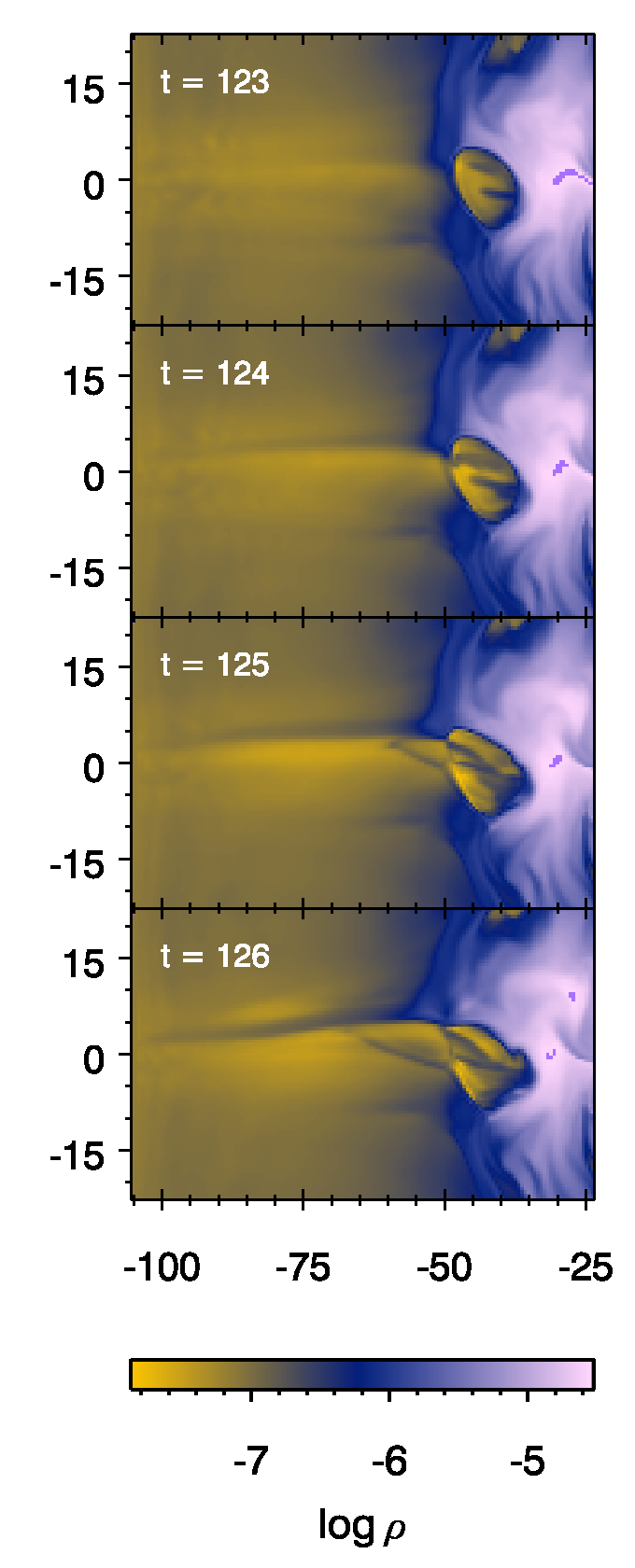}
      \caption{Density maps showing the evolution of the surge during its first minute on the $z=19$ plane. The axes show the box coordinates as displayed in Fig. \ref{density_t126_z19_2}.}
         \label{surge_evol}
   \end{figure}

\begin{figure}[h]
   \centering
   \includegraphics[scale=0.4]{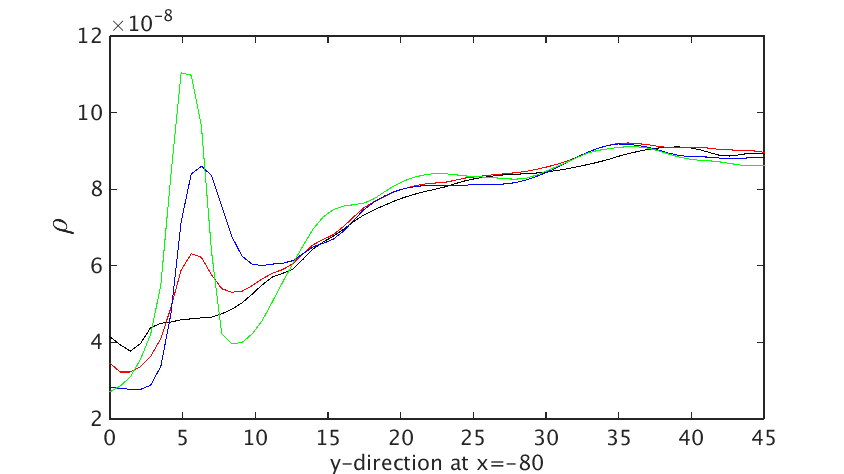}   
\caption{Cut in the $y$-direction through $x=-80$ at $z=19$ of density at different times. Key: black, $t=123$; red, $t=124$; blue, $t=125$; green, $t=126$. }
         \label{surge_den_slice}
   \end{figure}
.Figure \ref{surge_den_slice} shows the evolution of the density through a cut taken at $x=-80$. This is highlighted on  Fig. \ref{density_t126_z19_2} by a single line. The black curve in Fig \ref{surge_den_slice} shows the density profile through the cut just when the surge begins to flow at $t=123$. The other curves show the density through the cut at subsequent time steps (25 s apart). A distinct and growing peak develops with time between $y$=5 and $y$=10.  This corresponds to dense plasma flowing as a surge. Comparing these density profiles with the initial (black) curve, it is clear that the only significant change occurs in the thin region corresponding to the surge.

The overlying reconnection develops a very complex pattern that results in a strong eruption above the small EFR at $t=195$. Speeds of the ejected plasmoids are of the order of $\sim$40 ($\sim$272 km~s$^{-1}$). This phenomenon is mentioned here for completeness but is beyond the scope of the present work. The changing connectivities appear to be charateristic of recursive reconnection \citep{parnell08}.

\section{Asymmetric overlying field}
So far we have considered a symmetric flux emergence model and a symmetric model perturbed by a small EFR. \cite{guglielmino10} find that there is a directional bias, for the surges they observe, away from a nearby sunspot. The small EFR appears at one of the polarities of AR NOAA 10971. Hence one side is buffeted by (nearly) vertical magnetic field and the surges erupt in the opposite direction. We model this here by setting up a non-uniform initial ambient field that is vertical at one side of the computational domain and gradually becomes more horizontal as one moves to the opposite side. To achieve this, we take a two dimensional potential arcade of the form \citep[e.g.][]{dmac12}

\begin{eqnarray*}
B_x &=& B_p\cos[\pi(x-x_p)/{l}]e^{-\pi(z-z_p)/l}, \\
B_y &=& 0, \\
B_z &=& -B_p\sin[\pi(x-x_p)/{l}]e^{-\pi(z-z_p)/l},
\end{eqnarray*}
where $B_p$, $x_p$, $z_p$ and $l$ are constant parameters. For this ambient magnetic field, the side boundaries of the computational domain are now fixed and not periodic, as they were for the previous models. The boundary conditions in the $\pm y$ directions could still be set to periodic, however, we do not implement this here. To achieve the magnetic configuration described above, we take $B_p$ = 0.05 (65 G), $x_p$ = 100, $z_p$ = -25 and $l$ = 400. To start the simulation, a flux tube identical to that in the first model is inserted at $t=0$. The nondimensional background magnetic pressure, $P_m$, of the ambient magnetic field is a function of height, given by
\[
P_m({\rm background})= \frac{B_p^2}{2}e^{-2\pi(z-z_p)/l}.
\] 
This means there is no inherent horizontal gradient in the magnetic pressure of the ambient field.

As in the previous models, the magnitude of the magnetic field of the flux tube is significantly stronger than that of the ambient field at $t=0$. This is to allow the flux tube to rise to the photosphere relatively quickly. Once it undergoes the buoyancy instability and expands into the atmosphere, the effect of the asymmetric ambient field is clear. We shall only consider surges produced at the initial stage of emergence, corresponding to the Phase 1 surges of the  previous models. The development of stronger jets from different phases of reconnection, that can occur later, has been described in detail elsewhere \citep[e.g.][]{moreno13}. What we demonstrate here is how the choice of ambient field leads to surges in only one direction, as observed by \cite{guglielmino10}. Figure \ref{mag_field_t115} shows the magnetic field configuration at $t=115$.  Here, cyan field lines represent the ambient magnetic field and green field lines represent the EFR (connecting both footpoints). 

\begin{figure}[h]
   \centering
   \includegraphics[width=\hsize]{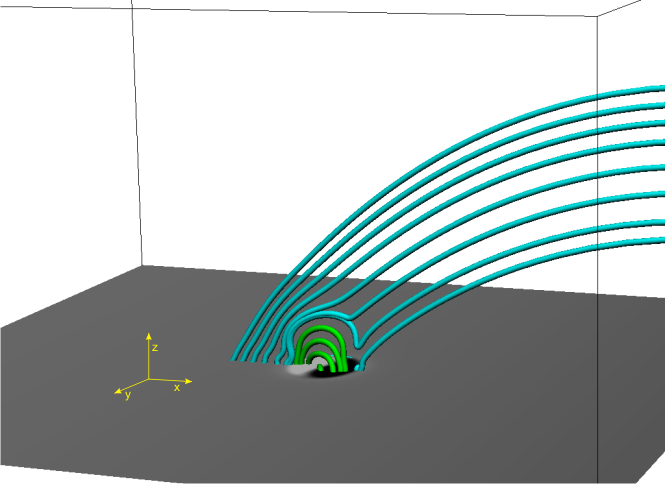}
      \caption{Selected magnetic field lines. These show how the emerging region reconnects with the asymmetric ambient magnetic field at $t=115$. The ambient field lines bend round the negative polarity to connect with the positive polarity or another position within the ambient field. A map of $B_z$ is placed at the base of the photosphere $(z=0)$.}
         \label{mag_field_t115}
   \end{figure}

On the positive side of the EFR (white footpoint on the vertical magnetic field map in Fig. \ref{mag_field_t115}) its magnetic field pushes into the ambient field and compresses it. On the other (negative) side, the geometry is conducive for reconnection. The form of reconnection found at this point in the evolution of the model is slightly different from the previous models. Ambient field lines enter the current sheet at the negative polarity of the EFR and change their connectivity. They do not reconnect to join the footpoints of the EFR but can change their positions within the ambient field. This is similar to `slipping reconnection' \citep[e.g.][]{priest95}, an entirely 3D process.

 When the EFR reaches the lower atmosphere, it dominates the magnetic pressure. Figure \ref{magp} displays a cut through the $y=0$ plane at height $z=10$ for $t=115$. 

\begin{figure}[h]
   \centering
   \includegraphics[width=\hsize]{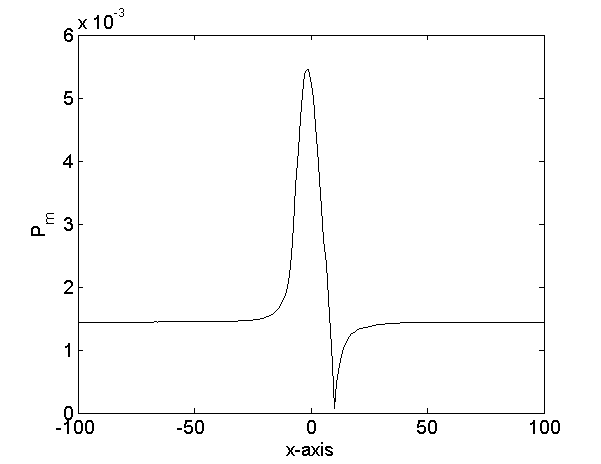}
      \caption{Magnetic pressure at $t=115$ at height $z=10$ in the $y=0$ plane. }
         \label{magp}
   \end{figure}

To the left of the central peak, the magnetic pressure decreases down to the background value for that height. To the right of the central peak, there is a dip in the magnetic pressure before it returns to the background value. This dip is due to reconnection on this side. Hence, even without an inherent horizontal magnetic pressure gradient in the background field, the geometry is enough to ensure that surges only flow on one side of the EFR.

Surges that develop emanate from the region of reduced magnetic pressure and flow along the ambient field away from the EFR. Figure \ref{s2} shows this, displaying a map of $(u_x^2+u_z^2)^{1/2}$ in the plane $y=0$ at $t=175$.

\begin{figure}[h]
   \centering
   \includegraphics[width=\hsize]{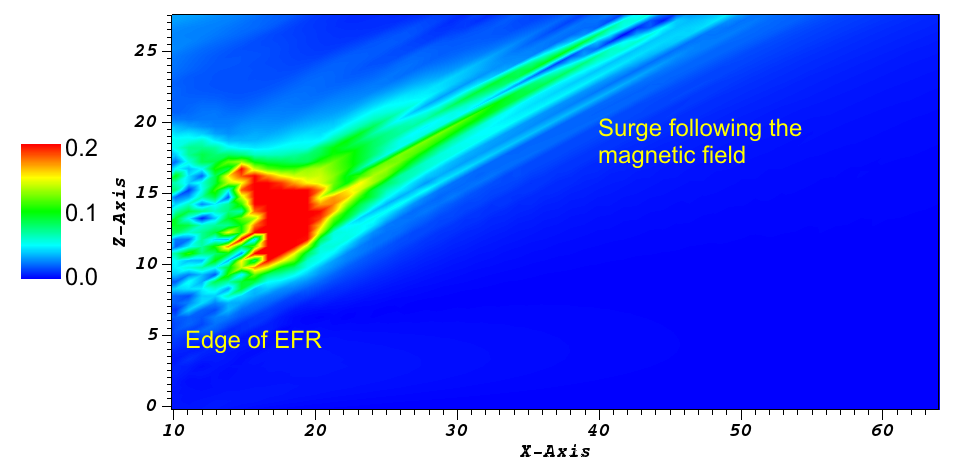}
      \caption{$(u_x^2+u_z^2)^{1/2}$ in the plane $y=0$ at $t=175$. The surge begins at the edge of the EFR and flows away from it, following the ambient magnetic field.}
         \label{s2}
   \end{figure}

\section{Discussion and conclusions}

In this paper we have presented three models to highlight the relationship between reconnection and surges during magnetic flux emergence. The models have been designed so that strong jets do not develop early on in the evolution. Instead, several phases of reconnection are found that correspond to the production of surges. When it comes to the evolution of surges, the first two models presented give a similar picture. In the first model, four phases of reconnection are identified. These are represented in the diagram of Fig. \ref{cartoon}  displayed earlier.  Phase 1 describes surges produced by reconnection at the sides of the EFR. Phase 2 sees (smooth) reconnection at a single point between the top of the EFR and the overlying magnetic field. Reconnection at the top of the EFR in the corona now connects to horizontal ambient field in the chromosphere. Hence, reconnection in the corona can precede the flow of surges in the chromosphere, as described in \cite{dominguez14}.   In Phase 3, a tearing-like instability occurs between the top of the EFR and the ambient field. This results in a more complex topology than in Phase 2. Finally, Phase 4 describes internal reconnection that can also produce surges.

The second model is the same as the first but with a smaller region emerging beside the main one to perturb it. Despite, however, the increased complexity in magnetic topology that another EFR brings, the behaviour of surge production is similar to that of the first model. That is, the general pattern of the reconnection phases still holds. 

The third model highlights how the geometry of the ambient field can influence the direction in which surges flow. \cite{guglielmino10} observe surges flowing in a preferred direction away from a sunspot. To mimic this, we choose an ambient field that is vertical at one side of the computational domain and gradually becomes horizontal as it reaches the other side. There is no inherent horizontal magnetic pressure gradient in the ambient field, so any preferred horizontal direction for surges is a result of interaction with the EFR. When the EFR reaches the atmosphere, reconnection is most efficient on one side and the surges generated have a preferred direction due to this.

Our models demonstrate the close connection between reconnection and surges in flux emergence. They can produce many features seen in observations, including (a) typical observed speeds; (b) evidence of reconnection in the corona generating surges that flow lower down in the chromosphere; (c) the filamentary density structure of surges; and (d) the manner in which the geometry of the ambient field determines the orientation of surges. In short, the models show how the evolving geometry of an emerging field, changed by distinct phases of reconnection, acts as a guide for the flowing of surges, which are themselves created by those same reconnection events. 

An interesting property of the models is that they follow a similar evolution to other studies where the reconnection is much stronger, such as in MH14 and \cite{moreno13}. The phases of reconnection follow a very similar pattern but can have different outcomes. In Phase 3 of the models presented here, although there are high flow speeds related to the development of a more complex magnetic topology, these quickly settle back down to typical surge values. In MH14, for example, there is also a phase of reconnection beginning with the tearing-like instability between the two flux systems after a phase of smooth reconnection. Due to the relative orientation of the two flux systems in MH14, however, reconnection is very efficient and produces strong jets. A large-scale (CME-type) eruption occurs in MH14 but not in the present models. This is simply due to the relative orientation between the EFR and the ambient field, which determines the efficacy of the overlying reconnection.  Phase 2 of the models can be detected observationally, as demonstrated by the modern observations of \cite{guglielmino10} and \cite{dominguez14}. Hence, so can Phase 3 and this may prove to be a key signature in helping to determine whether or not a region is likely to produce a large-scale (CME-type) eruption. 

One must be slightly cautious in generalizing the above idea, however. The flux emergence models described here are strictly for small-scale ARs.  It may be the case that larger active regions behave in a self-similar way. However, this remains to be tested. Also, if multiple regions emerge beside each other, as in the second model, high-speed (much higher than surge speeds) plasmoids can be ejected. However, there may be no subsequent CME-type eruption.

The results presented here are important for small-scale EFRs and represent an optimal benchmark for high resolution observations carried out by current and future telescopes, such as the GREGOR telescope \citep{schmidt12}, the European Solar Telescope \citep[EST, ][]{collados10}, and the Advanced Technology Solar Telescope \citep[ATST, ][]{keil10}.  The results may also prove to be important for space weather as such regions can erupt to produce CMEs \citep{mandrini05, schrijver10}.

\begin{acknowledgements}
DM, ALH and RS would like to thank the Carnegie Trust for financial support. DM, SLG and FZ were supported by the SOLARNET project (www.solarnet-east.eu), funded by the European Commision's FP7 Capacities Programme under the grant agreement no. 312495. RS was supported by Research Project Grant RPG-2012-600 funded by the Leverhulme Trust. FZ and SLG would like to acknowledge funding from the European Commission's Seventh Framework Programme under the grant agreements no. 284461 (eHEROES project) and no. 606862 (F-Chroma project). This research is also partly supported by the Italian MIUR-PRIN grant 2012P2HRCR on "The active Sun and its effects on Space and Earth climate" and by Space Weather Italian COmmunity (SWICO) Research Program. The computational work for this paper was carried out on the joint STFC and SFC (SRIF) funded cluster at the University of St Andrews. 
\end{acknowledgements}


\appendix
\section{Reconnection rate}

When the general theory of reconnection was developed, an elegant derivation of the reconnection rate was presented that makes use of an Euler potential representation of the magnetic field \citep{hesse88,schindler07}. Here we present an alternative derivation, based on Cartesian tensors, for resistive MHD. Consider magnetic field lines passing through a non-ideal region $\mathcal{D}$. Inside this region, the resistivity $\eta\ne 0$. Outside is an ideal plasma, where $\eta=0$. Consider a closed path that passes though $\mathcal{D}$ parallel to a magnetic field line, as shown in Fig. \ref{diffusion}
 \begin{figure}[h]
   \centering
   \includegraphics[scale=0.2]{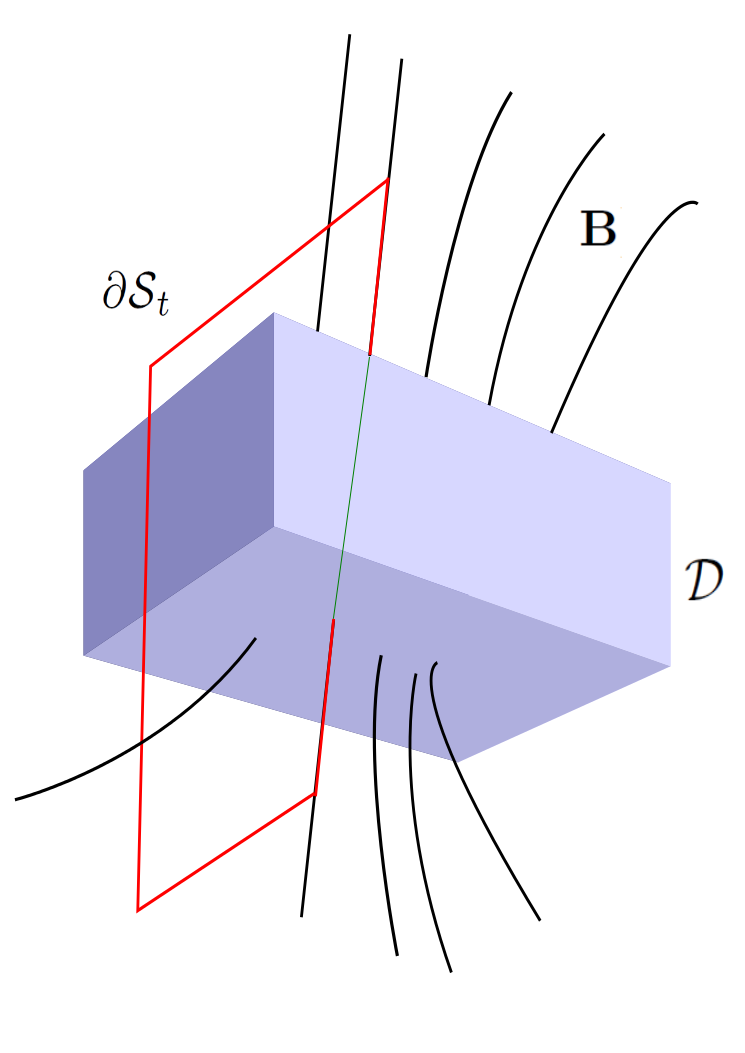}
      \caption{Magnetic field lines passing through a non-ideal region $\mathcal{D}$. A closed path, shown in red, defines the boundary of a surface $\mathcal{S}_t$. In $\mathcal{D}$, the path is parallel to the magnetic field. The section of the path inside $\mathcal{D}$ is shown in green. In the text, this is referred to as $\Gamma$.}
         \label{diffusion}
   \end{figure}

Let $\mathcal{S}_t$ denote the surface bounded by the closed path. The rate of reconnection is the rate of change of flux $\Phi$ through the surface
\[
\frac{{\rm d}\Phi}{{\rm d}t} = \frac{{\rm d}}{{\rm d}t}\int_{\mathcal{S}_t}\mathbf{B}\cdot\mathbf{n}\, {\rm d}a,
\]
where $\mathbf{n}$ is the unit normal to the surface $\mathcal{S}_t$. In order to differentiate the integral with respect to time, one must move from an Eulerian frame to a Lagrangian one. This can be achieved via an application of Nanson's formula to give
\begin{eqnarray*}
\frac{{\rm d}}{{\rm d}t}\int_{\mathcal{S}_t}\mathbf{B}\cdot\mathbf{n}\, {\rm d}a &=& \frac{{\rm d}}{{\rm d}t}\int_{\mathcal{S}_r}\mathbf{B}\cdot(J\mathbf{F}^{{-\rm T}}\mathbf{N})\,{\rm d}A, \\
&=& \int_{\mathcal{S}_r}\frac{{\rm d}}{{\rm d}t}(J\mathbf{F}^{-1}\mathbf{B})\cdot\mathbf{N}\,{\rm d}A, \\
&=& \int_{\mathcal{S}_r}\left[J\mathbf{F}^{-1}\dot{\mathbf{B}}+\dot{J}\mathbf{F}^{-1}\mathbf{B}+J\dot{{\mathbf{F}^{-1}}}\mathbf{B} \right]\cdot\mathbf{N}\,{\rm d} A.
\end{eqnarray*}
Here, geometric quantities that are now written in captials are relative to a Lagrangian frame with surface $\mathcal{S}_r$, e.g. $\mathbf{n}$ is Eulerian and $\mathbf{N}$ is Lagrangian. $\mathbf{F} = \partial\mathbf{x}/\partial\mathbf{X}$ is the deformation gradient, a second order Cartesian tensor relating the Lagrangian and Eulerian frames. Associated with this is $J={\rm det}(\mathbf{F})$. In the last integral, dots over terms represent differentiation with respect to time. It can be shown that
\[
\dot{J} = J{\rm tr}(\mathbf{L}),
\]
where $\mathbf{L} = \partial\mathbf{u}/\partial\mathbf{x}$. Here, $\mathbf{u}$ is the Eulerian velocity, making $\mathbf{L}$ an entirely Eulerian tensor. It can also be shown that
\[
\dot{{\mathbf{F}^{-1}}} = -\mathbf{F}^{-1}\mathbf{L},
\]
by differentiating the expression $\mathbf{F}\mathbf{F}^{-1} = \mathbf{I}$. Collecting these results together, it follows that
\begin{eqnarray*}
 \frac{{\rm d}}{{\rm d}t}\int_{\mathcal{S}_t}\mathbf{B}\cdot\mathbf{n}\, {\rm d}a &=& \int_{\mathcal{S}_r}[\dot{\mathbf{B}} + {\rm tr}(\mathbf{L})\mathbf{B}-\mathbf{L}\mathbf{B}]\cdot(J\mathbf{F}^{-{\rm T}}\mathbf{N})\,{\rm d}A, \\
&=& \int_{\mathcal{S}_t}[\dot{\mathbf{B}} + {\rm tr}(\mathbf{L})\mathbf{B}-\mathbf{L}\mathbf{B}]\cdot\mathbf{n}\,{\rm d}a.
\end{eqnarray*}
By expressing the integrand in terms of vectors and making use of the resistive induction equation, one finds
\begin{eqnarray*}
\dot{\mathbf{B}} + {\rm tr}(\mathbf{L})\mathbf{B}-\mathbf{L}\mathbf{B} &=& \frac{\partial\mathbf{B}}{\partial t} + (\mathbf{u}\cdot\nabla)\mathbf{B}+(\nabla\cdot\mathbf{u})\mathbf{B}-(\mathbf{B}\cdot\nabla)\mathbf{u}, \\
&=&  \frac{\partial\mathbf{B}}{\partial t} - \nabla\times(\mathbf{u}\times\mathbf{B}), \\
&=& -\nabla\times(\eta\nabla\times\mathbf{B}),
\end{eqnarray*}
where $\eta = \eta(\mathbf{x})$ is the resistivity. By an application of Stokes' theorem, one can show that
\begin{eqnarray*}
 \frac{{\rm d}}{{\rm d}t}\int_{\mathcal{S}_t}\mathbf{B}\cdot\mathbf{n}\, {\rm d}a &=& -\int_{\mathcal{S}_t}\nabla\times(\eta\nabla\times\mathbf{B})\cdot\mathbf{n}\,{\rm d}a, \\
&=& -\int_{\partial{\mathcal{S}_t}}\eta(\nabla\times\mathbf{B})\cdot\,{\rm d}\mathbf{l}.
\end{eqnarray*}
Taking the dot product of a unit vector $\hat{\mathbf{b}} = \mathbf{B}/|\mathbf{B}|$ with the resistive Ohm's law gives
\[
\mathbf{E}\cdot\hat{\mathbf{b}} + (\mathbf{u}\times\mathbf{B})\cdot\hat{\mathbf{b}} = E_{\|} = \sigma^{-1}\mathbf{j}\cdot\hat{\mathbf{b}} = \eta(\nabla\times\mathbf{B})\cdot\hat{\mathbf{b}},
\]
where $\eta=(\mu_0\sigma)^{-1}$ with conductivity $\sigma = \sigma(\mathbf{x})$ and (constant) magnetic permeability $\mu_0$. As $\eta = 0$ outside $\mathcal{D}$, the integration only gives a non-zero value along the section of the path inside $\mathcal{D}$ (the green path in Fig. \ref{diffusion}). If this section is labelled $\Gamma$, it follows that 
\begin{eqnarray*}
\int_{\partial\mathcal{S}_t}\eta(\nabla\times\mathbf{B})\cdot\,{\rm d}\mathbf{l} &=& \int_{\Gamma}\eta(\nabla\times\mathbf{B})\cdot\hat{\mathbf{b}}\,{\rm d}l, \\
&=& \int_{\Gamma}E_{\|}\,{\rm d}l.
\end{eqnarray*}
It then follows that
\[
\frac{{\rm d}}{{\rm d}t}\int_{\mathcal{S}_t}\mathbf{B}\cdot\mathbf{n}\, {\rm d}a = -\int_{\Gamma}E_{\|}\,{\rm d}l.
\] 
The choice of field line, and hence path, taken through $\mathcal{D}$ was arbitrary. Therefore, it is common to choose the field line that returns the largest magnitude. Since, in this work, we are integrating along field lines that pass through points connecting all four flux systems (four-colour points), the reconnection rate $\mathcal{R}(t)$ is taken to be
\[
\mathcal{R}(t) = \max_{\mathcal{N}(t)}\left|\int_{\Gamma}E_{\|}\,{\rm d}l\right|.
\]
Here the integration is along the field line that gives the largest magnitude for the integrated parallel electric field. $\mathcal{N}(t)$ is the set of field lines that pass through four-colour points, and varies in time. $\Gamma$ is the path along the field line in the current sheet.

\end{document}